\RequirePackage{lineno}
\documentclass[aps,prl,twocolumn,superscriptaddress,amsmath,amssymb,linenumbers]{revtex4}

\usepackage[colorlinks=true,breaklinks=true]{hyperref}
\hypersetup{allcolors=[rgb]{0.0 0.0 1.0},linkcolor=[rgb]{0.0 0.0 1.0}} 

\usepackage{graphicx} 
\usepackage{epstopdf} 
\usepackage{dcolumn} 
\usepackage{xcolor} 
\usepackage{amsmath,amssymb} 
\usepackage{hyperref} 
\usepackage[utf8]{inputenc} 
\usepackage[english]{babel} 
\usepackage{blindtext} 
\usepackage{ifthen} 
\usepackage{siunitx}
\usepackage{booktabs}
\usepackage{cleveref}
\usepackage{comment}
\usepackage{float}
\usepackage{afterpage}
\usepackage{wasysym}
\usepackage{booktabs}
\usepackage{makecell}
\usepackage[T1]{fontenc}
\AtBeginDocument{
\heavyrulewidth=.08em
\lightrulewidth=.05em
\cmidrulewidth=.03em
\belowrulesep=.65ex
\belowbottomsep=0pt
\aboverulesep=.4ex
\abovetopsep=0pt
\cmidrulesep=\doublerulesep
\cmidrulekern=.5em
\defaultaddspace=.5em
}

\graphicspath{{figures/}} 

\newboolean{articletitles}
\setboolean{articletitles}{true} 

\def\boosterpeakfirst {$\SI{18.55}{\GHz}$\xspace}
\def\boosterpeakfirstmass {$\SI{76.72}{\micro\eV}$\xspace}
\def\boosterpeaksecond {$\SI{19.21}{\GHz}$\xspace}
\def\boosterpeaksecondmass {$\SI{79.45}{\micro\eV}$\xspace}

\def\MADMAX       {\mbox{{\sc Madmax}}}  
\def\MADMAXbf     {\mbox{{\textbf{\textsc{Madmax}}}}}

\newboolean{wordcount}
\setboolean{wordcount}{false} 

\ifthenelse{\boolean{wordcount}}%
{\usepackage{bibentry} 
 \usepackage{comment} 
 \excludecomment{align*} 
 \excludecomment{align} 
 \excludecomment{equation*} 
 \excludecomment{equation} 
 \let\endequation\relax 
 \excludecomment{eqnarray*} 
 \excludecomment{eqnarray} 
 \let\endeqnarray\relax 
 \excludecomment{acknowledgments} 
 \def\maketitle{} 

      \makeatletter
      \newcommand*{\textalltt}{}
      \DeclareRobustCommand*{\textalltt}{%
	      \begingroup
	      \let\do\@makeother
	      \dospecials
	      \catcode`\\=\z@
	      \catcode`\{=\@ne
	      \catcode`\}=\tw@
	      \verbatim@font\@noligs
	      \@vobeyspaces
	      \frenchspacing
	      \@textalltt
      }
      \newcommand*{\@textalltt}[1]{%
	      #1%
	      \endgroup
      }
      \makeatother

}{}

\begin{document}


\title{First search for axion dark matter with a \MADMAXbf{} prototype
}

\begin{abstract}
This paper presents the first search for dark matter axions with mass in the ranges \SIrange{76.56}{76.82}{\micro\eV} and \SIrange{79.31}{79.53}{\micro\eV} using a prototype setup for the MAgnetized Disk and Mirror Axion eXperiment (\MADMAX{}). The experimental setup employs a dielectric haloscope consisting of three sapphire disks and a mirror to resonantly enhance 
the axion-induced microwave signal
within the magnetic dipole field provided by the \SI{1.6}{\tesla} Morpurgo magnet at CERN. Over 14.5 days of data collection, no axion signal was detected. 
A 95\% CL upper limit on the axion-photon coupling strength down to $|g_{a\gamma}| \sim 2 \times 10^{-11} \mathrm{GeV}^{-1}$ is set  in the targeted mass ranges, surpassing previous constraints, 
assuming a local axion dark matter density $\rho_{a}$ of $0.3~\mathrm{GeV}/\mathrm{cm}^3$. 
This study marks the first axion dark matter search using a dielectric haloscope.

\end{abstract}

\ifthenelse{\boolean{wordcount}}{}{
\newcommand{\aachen}{III. Physikalisches Institut A, RWTH Aachen University, Aachen,  Germany}
\newcommand{\mppb}{Max-Planck-Institut f\"{u}r Radioastronomie, Bonn, Germany}
\newcommand{\desy}{Deutsches Elektronen-Synchrotron DESY, Notkestr. 85, 22607 Hamburg, Germany}
\newcommand{\uhh}{Universit\"{a}t Hamburg, Hamburg, Germany}
\newcommand{\cppm}{Aix Marseille Université, CNRS/IN2P3, CPPM, Marseille, France}
\newcommand{\mppm}{Max-Planck-Institut f\"{u}r Physik, Garching, Germany}
\newcommand{\tubingen}{Physikalisches Institut, Eberhard Karls Universit\"{a}t T\"{u}bingen, T\"{u}bingen, Germany}
\newcommand{\zaragoza}{Universidad de Zaragoza, Zaragoza, Spain}
\newcommand{\ijclab}{Université Paris-Saclay, CNRS/IN2P3, IJCLab, Orsay, France}
\newcommand{\fnal}{Fermi National Accelerator Laboratory, Batavia, USA}

\author{\MADMAX{} Collaboration: B.~Ary dos Santos Garcia}
\affiliation{\aachen} 
\author{D.~Bergermann}
\affiliation{\aachen}
\author{A.~Caldwell}
\affiliation{\mppm}
\author{V.~Dabhi}
\affiliation{\cppm}
\author{C.~Diaconu}
\affiliation{\cppm}
\author{J.~Diehl}
\affiliation{\mppm}
\author{G.~Dvali}
\affiliation{\mppm}
\author{J.~Egge}
\affiliation{\uhh}
\author{E.~Garutti}
\affiliation{\uhh}
\author{S.~Heyminck}
\affiliation{\mppb}
\author{F.~Hubaut}
\affiliation{\cppm}
\author{A.~Ivanov}
\affiliation{\mppm}
\author{J.~Jochum}
\affiliation{\tubingen}
\author{S.~Knirck}
\affiliation{\fnal}
\author{M.~Kramer}
\affiliation{\mppb}
\author{D.~Kreikemeyer-Lorenzo}
\affiliation{\mppm}
\author{C.~Krieger}
\affiliation{\uhh}
\author{C.~Lee\footnote{Now at LLNL, Livermore, CA, USA}}
\affiliation{\mppm}
\author{D.~Leppla-Weber}
\affiliation{\desy}
\author{X.~Li\footnote{Now at TRIUMF, Vancouver, Canada}}
\affiliation{\mppm}
\author{A.~Lindner}
\affiliation{\desy}
\author{B.~Majorovits}
\affiliation{\mppm}
\author{J.P.A.~Maldonado}
\affiliation{\mppm}
\author{A.~Martini}
\affiliation{\desy}
\author{A.~Miyazaki}
\affiliation{\ijclab}
\author{E.~\"{O}z}
\affiliation{\aachen}
\author{P.~Pralavorio}
\affiliation{\cppm}
\author{G.~Raffelt}
\affiliation{\mppm}
\author{J.~Redondo}
\affiliation{\zaragoza}
\author{A.~Ringwald}
\affiliation{\desy}
\author{J.~Schaffran}
\affiliation{\desy}
\author{A.~Schmidt}
\affiliation{\aachen}
\author{F.~Steffen}
\affiliation{\mppm}
\author{C.~Strandhagen}
\affiliation{\tubingen}
\author{I.~Usherov}
\affiliation{\tubingen}
\author{H.~Wang}
\affiliation{\aachen}
\author{G.~Wieching}
\affiliation{\mppb}
}

\maketitle

\setlength\linenumbersep{5pt}
\textit{\textbf{Introduction}}---Axions have become prominent candidates for cold dark matter (DM)~\cite{Abbott:1982af,Dine:1982ah,Preskill:1982cy, Irastorza:2018dyq}. They were originally introduced to 
explain the absence of CP-violating effects in quantum chromodynamics (QCD) through the Peccei-Quinn (PQ) mechanism \cite{PhysRevLett.38.1440,PhysRevD.16.1791,PhysRevLett.40.223,PhysRevLett.40.279}. 
\begin{figure*}[ht!]
    \centering
    \includegraphics[width=0.80\textwidth]{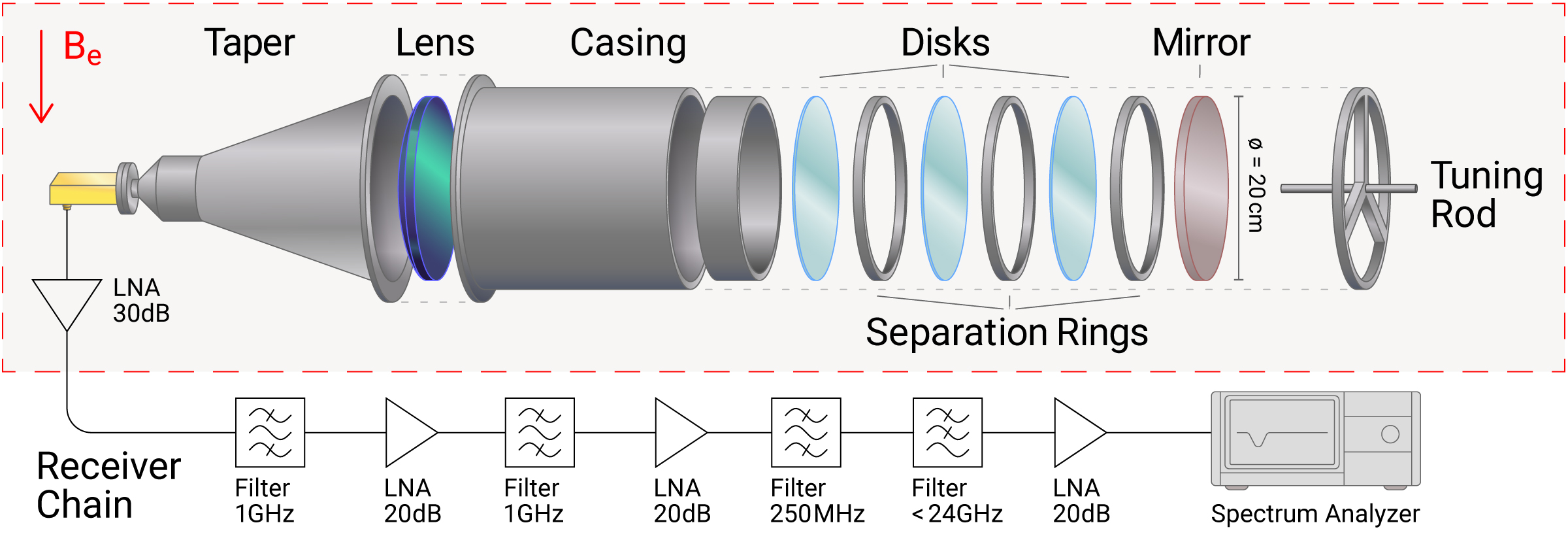}
    \caption{Exploded schematic view of the \MADMAX{} prototype CB200 and the receiver chain.  
    The  shaded region represents the components exposed to the B-field. }
    \label{fig:MADMAX}
\end{figure*}
In the well motivated post-inflationary PQ-symmetry breaking scenario the axion mass required  to match the observed DM density  
is expected to be in the range $m_a$\,$\sim$ \SIrange{26}{500}\,${\mu}$eV for axion models with short-lived domain walls \cite{Klaer:2017ond,Gorghetto:2020qws,Buschmann:2021sdq,Saikawa:2024bta} or even higher for other models~\cite{Ringwald:2015dsf, Beyer:2022ywc}. In the following, the term axion will refer to both, QCD axions and axion-like particles.

In an external magnetic field $\mathbf{B}_e$, axions convert to photons 
and source an oscillating current with frequency corresponding to $m_a$, given by $\mathbf{J}_a = g_{a\gamma} \mathbf{B}_e\dot a$ \cite{Millar_2017} as an additional term to be added to Maxwells' equations,  with $g_{a\gamma}$ the axion-photon coupling strength  and $\dot a$ the temporal derivative of the axion field.
Cavity searches are a promising approach to detect axions \cite{Sikivie:1983ip}. They are designed to convert axions from the DM halo into microwave photons using a resonator in a magnetic field.
Several cavity searches
~\cite{ADMX:2018gho, ADMX:2019uok, ADMX:2021nhd, Kim:2023vpo, CAPP:2024dtx, TASEH:2022vvu, Brubaker:2016ktl, HAYSTAC:2018rwy, HAYSTAC:2020kwv, Adair:2022rtw} 
have already put significant constraints on $|g_{a\gamma}|$ for $\SI{1}{\micro\eV}\lesssim~m_a~\lesssim~\SI{26}{\micro\eV}$, with first efforts being underway to test higher $m_a$~\cite{RADES:2020rlf, Quiskamp:2022pks, Quiskamp:2023ehr, QUAX:2023gop, QUAX:2024fut, Bajjali:2023uis, BREAD:2023xhc, Lawson:2019brd, CADEX, DeMiguel:2023nmz}, which are challenging to probe with cavity experiments as their size 
must be adjusted to match the wavelength of the generated photon. At higher $m_a$ values, corresponding to higher frequencies of the generated photon, this results in a reduced volume for resonant axion-photon conversion, thereby decreasing sensitivity.

The MAgnetized Disk and Mirror Axion eXperiment (\MADMAX{})~\cite{Jaeckel:2013eha,Caldwell:2016dcw, MADMAX:2019pub} makes use of the dielectric haloscope concept to access the well-motivated mass range $m_a > \SI{26}{\micro\eV}$. This concept, based on the magnetized mirror axion DM search idea~\cite{horns_2013}, uses a \textit{booster} composed of a set of parallel dielectric disks and a metallic mirror to enhance a potential axion signal, effectively decoupling the conversion volume from the wavelength and thereby overcoming the limitation of typical cavity experiments towards higher masses.

Due to the large de Broglie wavelength of the considered non-relativistic DM axions of $\mathcal{O}(\SI{10}{\m})$, $\mathbf{J}_a$ is assumed to be uniform over the extent of the booster. This leads to an oscillating electric field 
 uniform across a single medium with dielectric permittivity $\epsilon$.
The field $\mathbf{E}_a$ exhibits discontinuities at the boundaries of different media, resulting in the emission of traveling waves with frequency corresponding to $m_a$ ensuring the continuity of the electric field.
The signal power $P_\text{sig}$ due to the coherent emission from multiple surfaces in relation to the power emitted by an ideal magnetized mirror, $P_0$, 
is enhanced by the frequency-dependent boost factor $\beta^2 = P_\text{sig}/P_0$.
By adjusting the disk positions its center frequency and width can be tuned.
This makes it possible to boost the potential DM axion signal compared to the magnetized mirror concept~\cite{horns_2013} in dedicated mass ranges.
The sensitivity of a \MADMAX{} prototype setup to $|g_{a\gamma}|$ for a specific signal to noise ratio (SNR) can be estimated as \cite{Millar_2017}
\begin{equation}
    \label{eq1}
    \begin{split}
        |g_{a\gamma}| = &\;\SI{3.5e-11}{\giga\eV^{-1}} \sqrt{\frac{2\times 10^3}{{\beta^2}}} \sqrt{\frac{T_\mathrm{sys}}{\SI{300}{K}}} \\
        & \times\left( \frac{\SI{0.1}{m}}{r}\right) \left( \frac{\SI{1}{T}}{B_e} \right) \left( \frac{\SI{2.2}{days}}{\Delta t}\right)^{1/4} \!\!\sqrt{\frac{\mathrm{SNR}}{5}} \\
        & \times\left( \frac{m_a}{\SI{80}{\micro\eV}}\right)^{5/4} \sqrt{\frac{\SI{0.3}{\giga\eV/cm^3}}{\rho_a}}\;,
    \end{split}
\end{equation}
with system temperature $T_\mathrm{sys}$, radius $r$ of the disks and mirror, effective data-taking time $\Delta t$, and the local axion DM density~$\rho_a$.

The \MADMAX{} Collaboration has made significant progress in advancing the development of a future large-scale dielectric haloscope \cite{Knirck:2019eug,Egge:2020hyo,MADMAX:2021lxf,Egge:2022gfp,Egge:2023cos,macqu,Garutti:2023stk,MADMAX:2024pil}.
Small-scale setups have successfully validated the dielectric haloscope approach in dark photon DM searches \cite{Cervantes:2022epl, MADMAX:2024jnp}, specifically with a broadband \MADMAX{} prototype \cite{MADMAX:2024jnp}. 
This work for the first time demonstrates that a tunable dielectric haloscope can be used to search for axion DM.

\textit{\textbf{Experimental Setup}}---
The Closed Booster 200 (CB200) \MADMAX{} prototype is used.
As illustrated in \autoref{fig:MADMAX}, it consists of an aluminum mirror and three sapphire disks of $\SI{1}{\mm}$ thickness, each with a diameter of $\SI{200}{\mm}$ and distanced by separation rings enclosed in an aluminum casing, a Rexolite$^{\copyright}$ microwave lens and an aluminum taper, all controlled at $\mathcal{O}(\si{10\,\micro\m})$ precision.
In contrast to the design used in
\cite{MADMAX:2024jnp} and envisioned for the final \MADMAX{} design, the prototype is enclosed by conductive boundaries, resulting in fixed boundary conditions, such that only a limited number of cylindrical wave-guide modes within a given frequency range exists inside the booster.
This significantly simplifies modeling of the electromagnetic response and allows us to determine $\beta^2$ with fewer dedicated measurements.
Approximately 84\% of the axion induced power couples to the fundamental transverse electric TE$_{11}$ mode as calculated by the modal overlap formalism
considering the mode of propagation in a wave guide, which is absorbed in the calculation of $\beta^2$.
The emitted TE$_{11}$ power emissions are coupled into the receiver chain by a lens and a specifically designed taper  via a \SI{50}{\ohm} transmission line.
The power output of the system is coupled to a heterodyne receiver system, consisting of a series of low noise amplifiers (LNAs) and filters, and a \hbox{Rohde}~\&~Schwarz FSW43 real-time spectrum analyser (SA), which streams time-domain data with $\sim$\,0.11\,s coherent integration time to a computer where it is Fourier transformed on GPUs to power spectra. 
Subsequently, 8047 single power spectra are averaged to individual files with corresponding 8047\,$\times$\,0.11\,s\,=\,14.75\,min measurement time each.
The receiver system has a bandwidth of $\sim\SI{250}{\MHz}$, which is approximately centred around the respective $\beta^2$ peak, a resolution of $\SI{9}{\Hz}$ and provides negligible dead-time. It was configured for each run to acquire data around the frequency of the expected maximum $\beta^2$.
The gain of the  receiver chain  was calibrated by the Y-factor method ~\cite{pozar2011microwave} using a calibrated noise diode. 
$T_\mathrm{sys}$ is dominated by the $\sim$\,230\,K added noise of the
LNA.
Additional information on the setup and the methodology used can be found in \cite{supp}.

Two different disk spacing {\it configurations} leading to $\beta^2$ peaks in a $\sim$\,10\,MHz range  were used to search for axions: Configuration\,1 is designed to be sensitive around \boosterpeakfirst and configuration\,2 around \boosterpeaksecond, corresponding to axion masses around \boosterpeakfirstmass and \boosterpeaksecondmass, respectively.
Frequency adjustment is done by proper choice of the width of separation rings between adjacent disks and between mirror-facing disk and mirror: $\SI{12.52}{\mm}$, $\SI{12.25}{\mm}$ and $\SI{8.38}{\mm}$ (from left to right in \autoref{fig:MADMAX}) for configuration\,1 and $\SI{11.89}{\mm}$, $\SI{12.25}{\mm}$ and $\SI{8.02}{\mm}$ for configuration\,2.
The widths are optimized for a resonance of the TE$_{11}$ mode between mirror and adjacent disk. A $\mathcal{O}(\si{\micro\m})$ 
change in separation between
mirror and closest disk results in $\mathcal{O}(\si{\MHz})$ frequency shift of the $\beta^2$ distribution.
The peak frequencies for both configurations were precisely tuned for different runs to different values by applying mechanical force to the mirror using an adjustable tuning rod.
Three \textit{physics-runs} around \boosterpeakfirst and two around \boosterpeaksecond were performed, 
sensitive to different frequency ranges with peak frequencies separated by $\sim$\,10\,MHz.

Measurements of CB200 power spectra with a total measurement time of $14.5$\,days have been performed inside the Morpurgo dipole magnet \cite{Morpurgo} at CERN.
The magnet was operated at field strengths from $\SIrange{1\pm0.01}{1.58\pm0.02}{\tesla}$.

\textit{\textbf{Boost factor determination}}---The boost factor $\beta^{2}$, which is strongly correlated to the reflectivity S$_{11}$ of the system \cite{Millar_2017}, and its corresponding uncertainties are determined using a single-mode booster model \cite{supp} implemented in ADS \cite{keysight}.
The model is fitted to S$_{11}$ measurements of the booster. Disk positions, thicknesses, dielectric losses and permittivity are extracted. 
Losses due to the lens and three-dimensional effects, such as tilts, are modeled by effective dielectric losses.

Identification of the TE$_{11}$ booster mode is done for each configuration by measuring the shape of the E-field between mirror and closest disk, using a bead-pull method~\cite{Egge:2023cos}. It follows the expected shape.
S$_{11}$ measurements confirm the separation of the TE$_{11}$ modes from parasitic modes.

The $\beta^2$ calculation includes the overlap between the theoretical field shape and the uniform axion current of 84\%.
This is analogous to a form factor in cavity experiments.
The uncertainty on the overlap of $\pm 12\%$ is determined from the deviation between expected and measured field shape obtained from bead-pull measurements.

When measuring power spectra, the experimentally determined $\sim$\,41.5\,$\Omega$ impedance of the first LNA of the receiver chain leads to a mismatch with the transmission line
inducing a standing wave between booster and LNA. 
This modifies $\beta^2$ obtained from S$_{11}$ measurements, depending on the distance between the booster and this LNA. The distance is determined from the broadband oscillation pattern induced by resonance of the LNA noise, similar as in~\cite{ MADMAX:2024jnp}.
The modeled frequency behavior of the power spectra, adapted to consider the LNA impedance mismatch, matches the measured ones from the physics-runs.
Broadband power spectra measurements were taken prior to each physics-run. 
Drifts in $T_\mathrm{sys}$ or frequency are monitored by comparing the  individual 14.75\,min long measurements.
The variation of the measured booster peak frequencies during physics-runs is used to evaluate the systematic uncertainty on $\beta^{2}$.
They follow a Gaussian distribution with a width around  $\SI{200}{\kHz}$ for all physics-runs, resulting in an additional uncertainty in $\beta^{2}$ for a given frequency of less than $4\%$. 

\begin{figure}[b]
    \centering
    \includegraphics[width=0.99\linewidth]{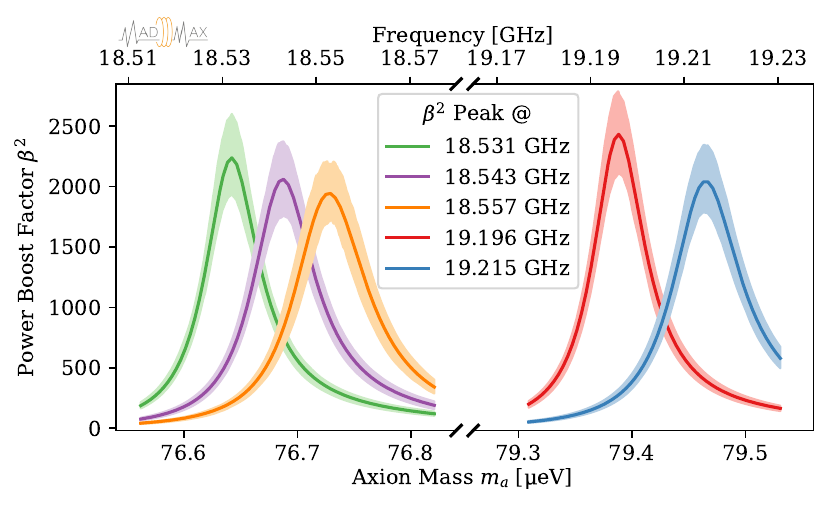}
    \caption{Modeled $\beta^2$ distributions for the five physics-runs. Lines
denote mean $\beta^2$ and shaded regions give $\pm 1 \sigma$ intervals.}
    \label{fig:boostfactors}
\end{figure}

The extracted boost factors including their uncertainties are shown in \autoref{fig:boostfactors}.
The maximum values for $\beta^2$ are around \SI{2000}, matching expectations~\cite{Millar_2017} and have uncertainties of 13 to 17\%. 

\begin{figure}[htbp]
    \centering
    \includegraphics[width=0.46\textwidth]{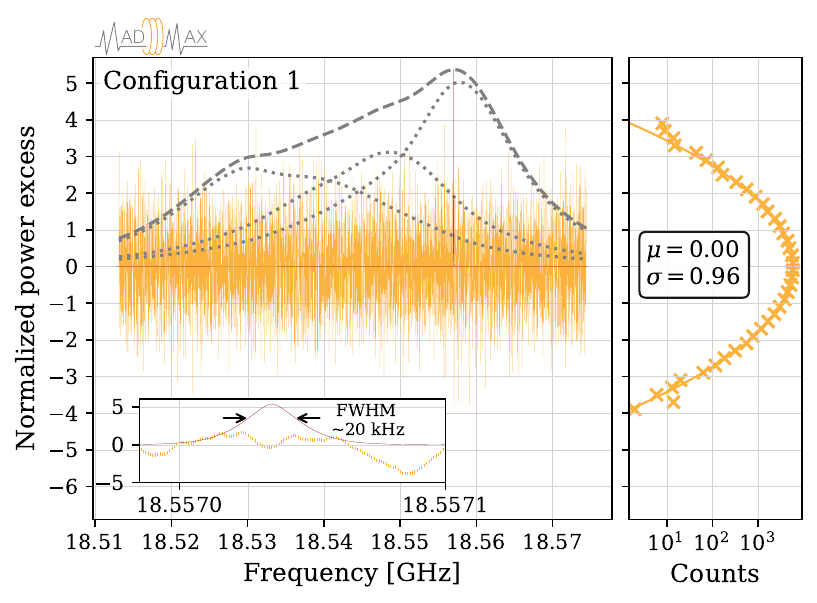}
    \includegraphics[width=0.46\textwidth]{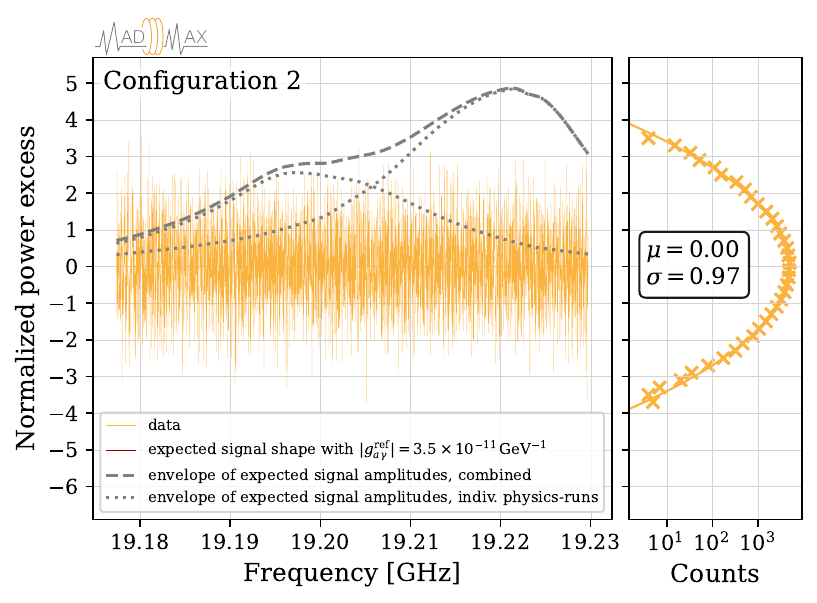}
    \caption{Grand spectra with 0.9\,kHz bin width for both booster configurations. Dashed gray lines indicate the envelope of expected amplitude in each frequency bin for a reference axion signal with photon coupling of $|g_{a \gamma}^\mathrm{ref}| = 3.5\times 10^{-11}\,$GeV$^{-1}$, combining all physics-runs for both configurations. Dotted lines show the expectation for individual physics runs. The expected signal shape of the same coupling strength at peak sensitivity is shown as solid line. The inset shows a zoom into this mass region.
    The right panels display histograms of the spectra, overlaid with a best-fit Gaussian distribution with the indicated mean and standard deviation.} 
    \label{fig:grand}
\end{figure}

\indent 
\textit{\textbf{Analysis}}---
The analysis procedure for the \MADMAX{} axion DM search \cite{johannes_thesis} is modeled after the framework established by HAYSTAC~\cite{Brubaker:2017rna} and adopts their nomenclature. It combines a set of individual power spectra $P_i(\nu)$, each measured with $\mathcal{O}(14.75\,\mathrm{min})$ averaging time, to a {\it grand spectrum}, where the full axion signal information is contained in a single bin. The raw FFT data are re-binned for analysis to $\SI{0.9}{\kHz}$, enough to resolve a potential ALP DM signal.

The spectra $P_i(\nu)$ are filtered using a fourth-order Savitzky-Golay (SG) filter~\cite{1964AnaCh..36.1627S} with a window length of $\sim \SI{1}{MHz}$ to extract the baselines $P_{\mathrm{bl},i}(\nu)$, which are subsequently used to obtain the processed spectra $p_{\mathrm{proc},i} = \left( P_i / P_{\mathrm{bl}, i} - 1\right)$. These processed spectra give local power excesses above the baseline across all frequency bins $i$. In the absence of an axion signal and assuming ideal baseline removal, the $p_{\mathrm{proc},i}$ are expected to follow Gaussian white noise with zero mean and a standard deviation determined by the averaging time. 
Studies of the full analysis chain applied to simulated data including synthetic axion signals reveal an SNR attenuation due to the SG filter by a factor of $\eta_\mathrm{SG} = 0.95$, which is constant in the range $1 < \mathrm{SNR} < 10$.

The baseline of the recorded power spectra
unexpectedly fluctuates on the scale of $\sim\SI{9}{\kHz}$ which introduces unwanted correlations between the processed spectra.
This sinusoidal fluctuation is removed by applying a digital notch filter 
on the inverse Fourier transforms of the $P_i(\nu)$. The filter has negligible influence on both amplitude and line-shape of any potential axion signal, while being effective at eliminating the 
fluctuation artifact. In combination with the SG filter, this process ensures uncorrelated $p_{\mathrm{proc},i}$.
The $p_{\mathrm{proc},i}$ are scaled and combined using weights that take the axion sensitivity depending on the B-field strength in individual runs into account.
The resulting combined spectrum is then cross-correlated with the expected Maxwell-Boltzmann axion line-shape~\cite{OHare:2017yze, Diehl:2023fuk} to yield the grand spectrum. 
A local DM velocity dispersion $\sigma_v = \SI{156}{km \per s}$~\cite{PhysRevD.42.3572} and a  velocity of the laboratory relative to the DM halo $v_\mathrm{lab} = \SI{242}{km \per s}$~\cite{2012ApJ...759..131B} are used, resulting in a FWHM of the expected line-shape of $\sim$\,20 kHz.

\begin{figure*}
    \centering
    \includegraphics[width=0.88\textwidth]{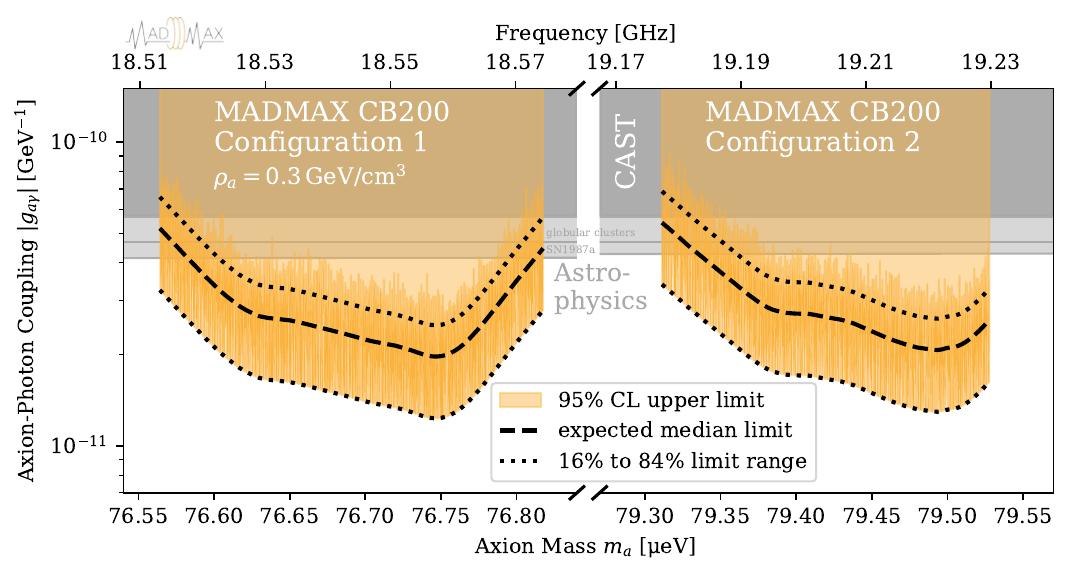}
    \caption{
    95\% CL exclusion limits in 0.9\,kHz bins (orange) from the 2024 \MADMAX{} axion search with the CB200 prototype using the Morpurgo magnet at CERN assuming a local axion DM density of $\rho_a = 0.3\,$GeV\,cm$^{-3}$. Limits are compared to the helioscope experiment CAST~\cite{CAST:2024eil} (dark grey) as well as to supernova \cite{PhysRevLett.133.211002} and globular cluster limits~\cite{Dolan:2022kul} (light grey). The expected median limit and the 16\% and 84\% quantiles are shown as black dashed and lower and upper dotted lines, respectively. The limits are truncated below the 16\% quantile.
    }
    \label{fig:limit}
\end{figure*}
The grand spectrum is unitless and, in the absence of an axion signal, expected from simulation to have zero mean and a standard deviation of 0.96, slightly below one due to correlations induced by the SG filter~\cite{Brubaker:2017rna}. Grand spectra for both configurations are presented in \autoref{fig:grand}.
The gray dashed line shows the envelope of the expected amplitude of excesses, reflecting the SNR of an axion signal with $|g_{a \gamma}^\mathrm{ref}| = 3.5 \times 10^{-11}\,$GeV$^{-1}$.
Its frequency dependence is determined by a combination of $T_\mathrm{sys}$ and $\beta^2$. 
An axion signal would appear as a narrow localized excess with FWHM of $\sim$\,20 kHz, corresponding to 22 bins at a frequency given by $m_a$ as shown in the inset.
The largest grand spectrum excess observed in both configurations has a local significance of $4.0\sigma$, which is consistent with statistical expectations. The probability of an equal or larger excess occurring across the entire dataset assuming no axion signal is $p=0.08$.

Since no evidence of DM axions is found, an upper limit on the axion-photon coupling $|g_{a \gamma}|$ is derived at 95\% Confidence Level (CL) for each 0.9\,kHz bin as shown in~\autoref{fig:limit}.
The individual uncertainty contributions on $|g_{a\gamma}|$ shown in~\autoref{tab:systematics}  are propagated into the total uncertainty taking correlations into account. For the $\beta^2$ determination this is done using using Monte Carlo simulation.

\begin{table}
\caption{Summary of systematic uncertainties on $\left\vert g_{a\gamma}\right\vert$. Minimum and maximum uncertainties among the frequency range for all physics-runs are reported 
as first and second values (when necessary). The uncertainty from each physics-run is considered individually when included in the limit estimation.}
\vspace{0.2cm}
\begin{tabular}{@{}lr@{}}
\toprule
Effect                                  & Uncertainty in $\left\vert g_{a\gamma}\right\vert$ \\ \midrule
Y-factor power calibration & $\SIrange{3}{5}{\percent}$ \\
Receiver chain power stability & $\leq \SI{2}{\percent}$\\ 
Axion  field -- TE$_{11}$ overlap & $\SI{6}{\percent}$\\
Boost factor determination (excl. overlap) & $< \SI{5}{\percent}$\\
Frequency stability of TE$_{11}$ mode & $< \SI{2}{\percent}$\\
\hline
Total & \SIrange{5}{10}{\percent} \\
\bottomrule
\end{tabular}
\label{tab:systematics} 
\end{table}

The resulting limits on DM axions exceed previous results from the CAST helioscope~\cite{CAST:2024eil} and astrophysical considerations~\cite{PhysRevLett.133.211002, Dolan:2022kul}.

\textit{\textbf{Conclusion}}---
Searching for axions with a dielectic haloscope, as proposed by the 
\MADMAX{} collaboration, facilitates accessing the “heavy” axion mass 
region favored by post-inflationary PQ-symmetry breaking scenarios.
A novel 
boost factor determination
method was developed and applied to derive first 
results from a prototype operated in CERN’s Morpurgo magnet, 
using the intrinsic advantages offered by the \MADMAX{} approach.
Detailed understanding of the  
radio frequency
response of
a booster system with closed boundary conditions was demonstrated.
The booster could be quickly tuned and re-calibrated 
to different frequencies demonstrating in principle the opportunity 
for future larger-range frequency scans.
The data from the prototype test campaign 
using a small booster system inside a modest B-field
allow probing for axion dark matter in previously 
uncharted territory.
The achievements presented here and in~\cite{MADMAX:2024jnp} provide 
a firm basis for the future research and development program towards 
a competitive dielectric haloscope, which will include the development 
of boosters with more and larger dielectric disks,
the demonstration of the calibration and operation of boosters at 
cryogenic temperatures and the implementation of tuning mechanisms 
via motorized disk position controls~\cite{Egge:2020hyo, Garutti:2023stk, MADMAX:2024pil}. 
With ongoing development of a future, unique, 
large-bore 9\,T dipole magnet~\cite{magnet_ieee, macqu}, 
the \MADMAX{} collaboration is on a very promising track towards 
probing dark matter QCD axions in the $100\,\si{\micro\eV}$ mass region.

\begin{acknowledgments}
{\color{black} \textit{\textbf{Acknowledgements}}---The authors would like to thank the CERN central cryogenic laboratory and the CERN magnet team for the support during all the measurements with the Morpurgo magnet.
We are thankful to A.~Kazemipour for helping setting up the experiment at CERN and for the RFI measurements performed during data taking. 
We thank O.~Reimann, D.~Strom and A.~Hambarzumjan for significant contributions during the first phase of the \MADMAX{} project. We thank O.~Rossel
for providing the drawing of the setup and artistic support. We acknowledge support by the Helmholtz Association (Germany); Deutsche Forschungsgemeinschaft (DFG, German Research Foundation) under Project No. 550641633 and Germany’s Excellence Strategy – EXC 2121 “Quantum Universe” – 390833306.
Computations were performed with computing resources granted by RWTH Aachen University under project 7936. SK is supported by Fermi Research Alliance, LLC under Contract No. DE-AC02-07CH11359 with the U.S. Department of Energy, Office of Science, Office of High Energy Physics.
We acknowledge the support of the \MADMAX{} project by the Max Planck Society.}

{\color{black} \textit{\textbf{Data availability}}---The data that support the findings of this letter are openly available \cite{open}}
\end{acknowledgments}

\bibliographystyle{JHEP}

\ifthenelse{\boolean{wordcount}}%
{ \nobibliography{references} }
{ \bibliography{references} }

\providecommand{\href}[2]{#2}\begingroup\raggedright\begin{thebibliography}{10}

\bibitem{Abbott:1982af}
L.~F. Abbott and P.~Sikivie, \emph{{A Cosmological Bound on the Invisible Axion}}, \href{https://doi.org/10.1016/0370-2693(83)90638-X}{\emph{Phys. Lett. B} {\bfseries 120} (1983) 133}.

\bibitem{Dine:1982ah}
M.~Dine and W.~Fischler, \emph{{The Not So Harmless Axion}}, \href{https://doi.org/10.1016/0370-2693(83)90639-1}{\emph{Phys. Lett. B} {\bfseries 120} (1983) 137}.

\bibitem{Preskill:1982cy}
J.~Preskill, M.~B. Wise and F.~Wilczek, \emph{{Cosmology of the Invisible Axion}}, \href{https://doi.org/10.1016/0370-2693(83)90637-8}{\emph{Phys. Lett. B} {\bfseries 120} (1983) 127}.

\bibitem{Irastorza:2018dyq}
I.~G. Irastorza and J.~Redondo, \emph{{New experimental approaches in the search for axion-like particles}}, \href{https://doi.org/10.1016/j.ppnp.2018.05.003}{\emph{Prog. Part. Nucl. Phys.} {\bfseries 102} (2018) 89} [\href{https://arxiv.org/abs/1801.08127}{{\ttfamily 1801.08127}}].

\bibitem{PhysRevLett.38.1440}
R.~D. Peccei and H.~R. Quinn, \emph{{CP Conservation in the Presence of Pseudoparticles}}, \href{https://doi.org/10.1103/PhysRevLett.38.1440}{\emph{Phys. Rev. Lett.} {\bfseries 38} (1977) 1440}.

\bibitem{PhysRevD.16.1791}
R.~D. Peccei and H.~R. Quinn, \emph{{Constraints imposed by CP conservation in the presence of pseudoparticles}}, \href{https://doi.org/10.1103/PhysRevD.16.1791}{\emph{Phys. Rev. D} {\bfseries 16} (1977) 1791}.

\bibitem{PhysRevLett.40.223}
S.~Weinberg, \emph{A new light boson?}, \href{https://doi.org/10.1103/PhysRevLett.40.223}{\emph{Phys. Rev. Lett.} {\bfseries 40} (1978) 223}.

\bibitem{PhysRevLett.40.279}
F.~Wilczek, \emph{{Problem of Strong $P$ and $T$ Invariance in the Presence of Instantons}}, \href{https://doi.org/10.1103/PhysRevLett.40.279}{\emph{Phys. Rev. Lett.} {\bfseries 40} (1978) 279}.

\bibitem{Klaer:2017ond}
V.~B.~. Klaer and G.~D. Moore, \emph{{The dark-matter axion mass}}, \href{https://doi.org/10.1088/1475-7516/2017/11/049}{\emph{JCAP} {\bfseries 11} (2017) 049} [\href{https://arxiv.org/abs/1708.07521}{{\ttfamily 1708.07521}}].

\bibitem{Gorghetto:2020qws}
M.~Gorghetto, E.~Hardy and G.~Villadoro, \emph{{More axions from strings}}, \href{https://doi.org/10.21468/SciPostPhys.10.2.050}{\emph{SciPost Phys.} {\bfseries 10} (2021) 050} [\href{https://arxiv.org/abs/2007.04990}{{\ttfamily 2007.04990}}].

\bibitem{Buschmann:2021sdq}
M.~Buschmann, J.~W. Foster, A.~Hook, A.~Peterson, D.~E. Willcox, W.~Zhang et~al., \emph{{Dark matter from axion strings with adaptive mesh refinement}}, \href{https://doi.org/10.1038/s41467-022-28669-y}{\emph{Nature Commun.} {\bfseries 13} (2022) 1049} [\href{https://arxiv.org/abs/2108.05368}{{\ttfamily 2108.05368}}].

\bibitem{Saikawa:2024bta}
K.~Saikawa, J.~Redondo, A.~Vaquero and M.~Kaltschmidt, \emph{{Spectrum of global string networks and the axion dark matter mass}}, \href{https://doi.org/10.1088/1475-7516/2024/10/043}{\emph{JCAP} {\bfseries 10} (2024) 043} [\href{https://arxiv.org/abs/2401.17253}{{\ttfamily 2401.17253}}].

\bibitem{Ringwald:2015dsf}
A.~Ringwald and K.~Saikawa, \emph{{Axion dark matter in the post-inflationary Peccei-Quinn symmetry breaking scenario}}, \href{https://doi.org/10.1103/PhysRevD.93.085031}{\emph{Phys. Rev. D} {\bfseries 93} (2016) 085031} [\href{https://arxiv.org/abs/1512.06436}{{\ttfamily 1512.06436}}].

\bibitem{Beyer:2022ywc}
K.~A. Beyer and S.~Sarkar, \emph{{Ruling out light axions: The writing is on the wall}}, \href{https://doi.org/10.21468/SciPostPhys.15.1.003}{\emph{SciPost Phys.} {\bfseries 15} (2023) 003} [\href{https://arxiv.org/abs/2211.14635}{{\ttfamily 2211.14635}}].

\bibitem{Millar_2017}
A.~J. Millar, G.~G. Raffelt, J.~Redondo and F.~D. Steffen, \emph{{Dielectric Haloscopes to Search for Axion Dark Matter: Theoretical Foundations}}, \href{https://doi.org/10.1088/1475-7516/2017/01/061}{\emph{JCAP} {\bfseries 01} (2017) 061} [\href{https://arxiv.org/abs/1612.07057}{{\ttfamily 1612.07057}}].

\bibitem{Sikivie:1983ip}
P.~Sikivie, \emph{{Experimental Tests of the Invisible Axion}}, \href{https://doi.org/10.1103/PhysRevLett.51.1415}{\emph{Phys. Rev. Lett.} {\bfseries 51} (1983) 1415}.

\bibitem{ADMX:2018gho}
{\scshape ADMX} collaboration, N.~Du et~al., \emph{{A Search for Invisible Axion Dark Matter with the Axion Dark Matter Experiment}}, \href{https://doi.org/10.1103/PhysRevLett.120.151301}{\emph{Phys. Rev. Lett.} {\bfseries 120} (2018) 151301} [\href{https://arxiv.org/abs/1804.05750}{{\ttfamily 1804.05750}}].

\bibitem{ADMX:2019uok}
{\scshape ADMX} collaboration, T.~Braine et~al., \emph{{Extended Search for the Invisible Axion with the Axion Dark Matter Experiment}}, \href{https://doi.org/10.1103/PhysRevLett.124.101303}{\emph{Phys. Rev. Lett.} {\bfseries 124} (2020) 101303} [\href{https://arxiv.org/abs/1910.08638}{{\ttfamily 1910.08638}}].

\bibitem{ADMX:2021nhd}
{\scshape ADMX} collaboration, C.~Bartram et~al., \emph{{Search for Invisible Axion Dark Matter in the 3.3\textendash{}4.2\,\,\ensuremath{\mu}eV Mass Range}}, \href{https://doi.org/10.1103/PhysRevLett.127.261803}{\emph{Phys. Rev. Lett.} {\bfseries 127} (2021) 261803} [\href{https://arxiv.org/abs/2110.06096}{{\ttfamily 2110.06096}}].

\bibitem{Kim:2023vpo}
Y.~Kim et~al., \emph{{Experimental Search for Invisible Dark Matter Axions around 22\,\,\ensuremath{\mu}eV}}, \href{https://doi.org/10.1103/PhysRevLett.133.051802}{\emph{Phys. Rev. Lett.} {\bfseries 133} (2024) 051802} [\href{https://arxiv.org/abs/2312.11003}{{\ttfamily 2312.11003}}].

\bibitem{CAPP:2024dtx}
{\scshape CAPP} collaboration, S.~Ahn et~al., \emph{{Extensive Search for Axion Dark Matter over 1~GHz with CAPP\textquoteright{}S Main Axion Experiment}}, \href{https://doi.org/10.1103/PhysRevX.14.031023}{\emph{Phys. Rev. X} {\bfseries 14} (2024) 031023} [\href{https://arxiv.org/abs/2402.12892}{{\ttfamily 2402.12892}}].

\bibitem{TASEH:2022vvu}
{\scshape TASEH} collaboration, H.~Chang et~al., \emph{{First Results from the Taiwan Axion Search Experiment with a Haloscope at 19.6\,\,\ensuremath{\mu}eV}}, \href{https://doi.org/10.1103/PhysRevLett.129.111802}{\emph{Phys. Rev. Lett.} {\bfseries 129} (2022) 111802} [\href{https://arxiv.org/abs/2205.05574}{{\ttfamily 2205.05574}}].

\bibitem{Brubaker:2016ktl}
B.~M. Brubaker et~al., \emph{{First results from a microwave cavity axion search at 24 $\mu$eV}}, \href{https://doi.org/10.1103/PhysRevLett.118.061302}{\emph{Phys. Rev. Lett.} {\bfseries 118} (2017) 061302} [\href{https://arxiv.org/abs/1610.02580}{{\ttfamily 1610.02580}}].

\bibitem{HAYSTAC:2018rwy}
{\scshape HAYSTAC} collaboration, L.~Zhong et~al., \emph{{Results from phase 1 of the HAYSTAC microwave cavity axion experiment}}, \href{https://doi.org/10.1103/PhysRevD.97.092001}{\emph{Phys. Rev. D} {\bfseries 97} (2018) 092001} [\href{https://arxiv.org/abs/1803.03690}{{\ttfamily 1803.03690}}].

\bibitem{HAYSTAC:2020kwv}
{\scshape HAYSTAC} collaboration, K.~M. Backes et~al., \emph{{A quantum-enhanced search for dark matter axions}}, \href{https://doi.org/10.1038/s41586-021-03226-7}{\emph{Nature} {\bfseries 590} (2021) 238} [\href{https://arxiv.org/abs/2008.01853}{{\ttfamily 2008.01853}}].

\bibitem{Adair:2022rtw}
C.~M. Adair et~al., \emph{{Search for Dark Matter Axions with CAST-CAPP}}, \href{https://doi.org/10.1038/s41467-022-33913-6}{\emph{Nature Commun.} {\bfseries 13} (2022) 6180} [\href{https://arxiv.org/abs/2211.02902}{{\ttfamily 2211.02902}}].

\bibitem{RADES:2020rlf}
{\scshape RADES} collaboration, A.~A. Melc\'on et~al., \emph{{First results of the CAST-RADES haloscope search for axions at 34.67 $\mu$eV}}, \href{https://doi.org/10.1007/JHEP10(2021)075}{\emph{JHEP} {\bfseries 21} (2020) 075} [\href{https://arxiv.org/abs/2104.13798}{{\ttfamily 2104.13798}}].

\bibitem{Quiskamp:2022pks}
A.~P. Quiskamp, B.~T. McAllister, P.~Altin, E.~N. Ivanov, M.~Goryachev and M.~E. Tobar, \emph{{Direct search for dark matter axions excluding ALP cogenesis in the 63- to 67-\ensuremath{\mu}eV range with the ORGAN experiment}}, \href{https://doi.org/10.1126/sciadv.abq3765}{\emph{Sci. Adv.} {\bfseries 8} (2022) abq3765} [\href{https://arxiv.org/abs/2203.12152}{{\ttfamily 2203.12152}}].

\bibitem{Quiskamp:2023ehr}
A.~Quiskamp, B.~T. McAllister, P.~Altin, E.~N. Ivanov, M.~Goryachev and M.~E. Tobar, \emph{{Exclusion of Axionlike-Particle Cogenesis Dark Matter in a Mass Window above 100\,\,\ensuremath{\mu}eV}}, \href{https://doi.org/10.1103/PhysRevLett.132.031601}{\emph{Phys. Rev. Lett.} {\bfseries 132} (2024) 031601} [\href{https://arxiv.org/abs/2310.00904}{{\ttfamily 2310.00904}}].

\bibitem{QUAX:2023gop}
{\scshape QUAX} collaboration, R.~Di~Vora et~al., \emph{{Search for galactic axions with a traveling wave parametric amplifier}}, \href{https://doi.org/10.1103/PhysRevD.108.062005}{\emph{Phys. Rev. D} {\bfseries 108} (2023) 062005} [\href{https://arxiv.org/abs/2304.07505}{{\ttfamily 2304.07505}}].

\bibitem{QUAX:2024fut}
{\scshape QUAX} collaboration, A.~Rettaroli et~al., \emph{{Search for axion dark matter with the QUAX\textendash{}LNF tunable haloscope}}, \href{https://doi.org/10.1103/PhysRevD.110.022008}{\emph{Phys. Rev. D} {\bfseries 110} (2024) 022008} [\href{https://arxiv.org/abs/2402.19063}{{\ttfamily 2402.19063}}].

\bibitem{Bajjali:2023uis}
F.~Bajjali et~al., \emph{{First results from BRASS-p broadband searches for hidden photon dark matter}}, \href{https://doi.org/10.1088/1475-7516/2023/08/077}{\emph{JCAP} {\bfseries 08} (2023) 077} [\href{https://arxiv.org/abs/2306.05934}{{\ttfamily 2306.05934}}].

\bibitem{BREAD:2023xhc}
{\scshape BREAD} collaboration, S.~Knirck et~al., \emph{{First Results from a Broadband Search for Dark Photon Dark Matter in the 44 to 52\,\,\ensuremath{\mu}eV Range with a Coaxial Dish Antenna}}, \href{https://doi.org/10.1103/PhysRevLett.132.131004}{\emph{Phys. Rev. Lett.} {\bfseries 132} (2024) 131004} [\href{https://arxiv.org/abs/2310.13891}{{\ttfamily 2310.13891}}].

\bibitem{Lawson:2019brd}
M.~Lawson, A.~J. Millar, M.~Pancaldi, E.~Vitagliano and F.~Wilczek, \emph{{Tunable axion plasma haloscopes}}, \href{https://doi.org/10.1103/PhysRevLett.123.141802}{\emph{Phys. Rev. Lett.} {\bfseries 123} (2019) 141802} [\href{https://arxiv.org/abs/1904.11872}{{\ttfamily 1904.11872}}].

\bibitem{CADEX}
B.~Aja et~al., \emph{{The Canfranc Axion Detection Experiment (CADEx): search for axions at 90 GHz with Kinetic Inductance Detectors}}, \href{https://doi.org/10.1088/1475-7516/2022/11/044}{\emph{JCAP} {\bfseries 11} (2022) 044} [\href{https://arxiv.org/abs/2206.02980}{{\ttfamily 2206.02980}}].

\bibitem{DeMiguel:2023nmz}
{\scshape DALI} collaboration, J.~De~Miguel, J.~F. Hern\'andez-Cabrera, E.~Hern\'andez-Su\'arez, E.~Joven-\'Alvarez, C.~Otani and J.~A. Rubi\~no Mart\'\i{}n, \emph{{Discovery prospects with the Dark-photons \& Axion-like particles Interferometer}}, \href{https://doi.org/10.1103/PhysRevD.109.062002}{\emph{Phys. Rev. D} {\bfseries 109} (2024) 062002} [\href{https://arxiv.org/abs/2303.03997}{{\ttfamily 2303.03997}}].

\bibitem{Jaeckel:2013eha}
J.~Jaeckel and J.~Redondo, \emph{{Resonant to broadband searches for cold dark matter consisting of weakly interacting slim particles}}, \href{https://doi.org/10.1103/PhysRevD.88.115002}{\emph{Phys. Rev. D} {\bfseries 88} (2013) 115002} [\href{https://arxiv.org/abs/1308.1103}{{\ttfamily 1308.1103}}].

\bibitem{Caldwell:2016dcw}
A.~Caldwell, G.~Dvali, B.~Majorovits, A.~Millar, G.~Raffelt, J.~Redondo et~al., \emph{{Dielectric Haloscopes: A New Way to Detect Axion Dark Matter}}, \href{https://doi.org/10.1103/PhysRevLett.118.091801}{\emph{Phys. Rev. Lett.} {\bfseries 118} (2017) 091801} [\href{https://arxiv.org/abs/1611.05865}{{\ttfamily 1611.05865}}].

\bibitem{MADMAX:2019pub}
{\scshape \MADMAX} collaboration, P.~Brun et~al., \emph{{A new experimental approach to probe QCD axion dark matter in the mass range above 40 $\mu$eV}}, \href{https://doi.org/10.1140/epjc/s10052-019-6683-x}{\emph{Eur. Phys. J. C} {\bfseries 79} (2019) 186} [\href{https://arxiv.org/abs/1901.07401}{{\ttfamily 1901.07401}}].

\bibitem{horns_2013}
D.~Horns, J.~Jaeckel, A.~Lindner, A.~Lobanov, J.~Redondo and A.~Ringwald, \emph{{Searching for WISPy Cold Dark Matter with a Dish Antenna}}, \href{https://doi.org/10.1088/1475-7516/2013/04/016}{\emph{JCAP} {\bfseries 04} (2013) 016} [\href{https://arxiv.org/abs/1212.2970}{{\ttfamily 1212.2970}}].

\bibitem{Knirck:2019eug}
S.~Knirck, J.~Sch\"utte-Engel, A.~Millar, J.~Redondo, O.~Reimann, A.~Ringwald et~al., \emph{{A First Look on 3D Effects in Open Axion Haloscopes}}, \href{https://doi.org/10.1088/1475-7516/2019/08/026}{\emph{JCAP} {\bfseries 08} (2019) 026} [\href{https://arxiv.org/abs/1906.02677}{{\ttfamily 1906.02677}}].

\bibitem{Egge:2020hyo}
J.~Egge, S.~Knirck, B.~Majorovits, C.~Moore and O.~Reimann, \emph{{A first proof of principle booster setup for the \MADMAX{} dielectric haloscope}}, \href{https://doi.org/10.1140/epjc/s10052-020-7985-8}{\emph{Eur. Phys. J. C} {\bfseries 80} (2020) 392} [\href{https://arxiv.org/abs/2001.04363}{{\ttfamily 2001.04363}}].

\bibitem{MADMAX:2021lxf}
{\scshape \MADMAX} collaboration, S.~Knirck et~al., \emph{{Simulating \MADMAX{} in 3D: requirements for dielectric axion haloscopes}}, \href{https://doi.org/10.1088/1475-7516/2021/10/034}{\emph{JCAP} {\bfseries 10} (2021) 034} [\href{https://arxiv.org/abs/2104.06553}{{\ttfamily 2104.06553}}].

\bibitem{Egge:2022gfp}
J.~Egge, \emph{{Axion haloscope signal power from reciprocity}}, \href{https://doi.org/10.1088/1475-7516/2023/04/064}{\emph{JCAP} {\bfseries 04} (2023) 064} [\href{https://arxiv.org/abs/2211.11503}{{\ttfamily 2211.11503}}].

\bibitem{Egge:2023cos}
J.~Egge et~al., \emph{{Experimental determination of axion signal power of dish antennas and dielectric haloscopes using the reciprocity approach}}, \href{https://doi.org/10.1088/1475-7516/2024/04/005}{\emph{JCAP} {\bfseries 04} (2024) 005} [\href{https://arxiv.org/abs/2311.13359}{{\ttfamily 2311.13359}}].

\bibitem{macqu}
C.~Lorin, W.~A. Maksoud, J.~Allard, C.~Berriaud, V.~Calvelli, L.~Denarie et~al., \emph{{Development, Integration, and Test of the MACQU Demo Coil Toward \MADMAX{} Quench Analysis}}, \href{https://doi.org/10.1109/TASC.2023.3273734}{\emph{IEEE TAS} {\bfseries 33} (2023) 1}.

\bibitem{Garutti:2023stk}
E.~Garutti, H.~Janssen, D.~Kreikemeyer-Lorenzo, C.~Krieger, A.~Lindner, B.~Majorovits et~al., \emph{{Qualification of piezo-electric actuators for the \MADMAX{} booster system at cryogenic temperatures and high magnetic fields}}, \href{https://doi.org/10.1088/1748-0221/18/08/P08011}{\emph{JINST} {\bfseries 18} (2023) P08011} [\href{https://arxiv.org/abs/2305.12808}{{\ttfamily 2305.12808}}].

\bibitem{MADMAX:2024pil}
{\scshape \MADMAX} collaboration, B.~Ary Dos Santos~Garcia et~al., \emph{{First mechanical realization of a tunable dielectric haloscope for the \MADMAX{} axion search experiment}}, \href{https://doi.org/10.1088/1748-0221/19/11/T11002}{\emph{JINST} {\bfseries 19} (2024) T11002} [\href{https://arxiv.org/abs/2407.10716}{{\ttfamily 2407.10716}}].

\bibitem{Cervantes:2022epl}
R.~Cervantes et~al., \emph{{ADMX-Orpheus first search for 70\,\,\ensuremath{\mu}eV dark photon dark matter: Detailed design, operations, and analysis}}, \href{https://doi.org/10.1103/PhysRevD.106.102002}{\emph{Phys. Rev. D} {\bfseries 106} (2022) 102002} [\href{https://arxiv.org/abs/2204.09475}{{\ttfamily 2204.09475}}].

\bibitem{MADMAX:2024jnp}
{\scshape \MADMAX} collaboration, J.~Egge et~al., \emph{{First Search for Dark Photon Dark Matter with a \MADMAX{} Prototype}}, \href{https://doi.org/10.1103/PhysRevLett.134.151004}{\emph{Phys. Rev. Lett.} {\bfseries 134} (2025) 151004} [\href{https://arxiv.org/abs/2408.02368}{{\ttfamily 2408.02368}}].

\bibitem{pozar2011microwave}
D.~Pozar, \emph{Microwave Engineering}. Wiley, 2011.

\bibitem{supp}
See supplemental material at [URL] for additional information on the overlap formalism, preparation and verification of the booster mode, the experimental setup used at CERN, data taking, power calibration and booster noise modelling. The supplemental material also contains \cite{4065916}.

\bibitem{Morpurgo}
M.~Morpurgo, \emph{A large superconducting dipole cooled by forced circulation of two phase helium}, \href{https://doi.org/10.1016/0011-2275(79)90126-7}{\emph{Cryogenics} {\bfseries 19} (1979) 411}.

\bibitem{keysight}
Keysight, ``{ADS} advanced design system.'' \url{https://www.keysight.com/de/de/products/software/pathwave-design-software/pathwave-advanced-design-system.html}.

\bibitem{johannes_thesis}
J.~Diehl, \emph{Statistical Methods for a First \MADMAX{} Axion Dark Matter Search and Beyond}, Ph.D. thesis, Technical University Munich, 2024, Available at \url{https://mediatum.ub.tum.de/?id=1752277}.

\bibitem{Brubaker:2017rna}
B.~M. Brubaker, L.~Zhong, S.~K. Lamoreaux, K.~W. Lehnert and K.~A. van Bibber, \emph{{HAYSTAC axion search analysis procedure}}, \href{https://doi.org/10.1103/PhysRevD.96.123008}{\emph{Phys. Rev. D} {\bfseries 96} (2017) 123008} [\href{https://arxiv.org/abs/1706.08388}{{\ttfamily 1706.08388}}].

\bibitem{1964AnaCh..36.1627S}
A.~{Savitzky} and M.~J.~E. {Golay}, \emph{{Smoothing and differentiation of data by simplified least squares procedures}}, \href{https://doi.org/10.1021/ac60214a047}{\emph{Analytical Chemistry} {\bfseries 36} (1964) 1627}.

\bibitem{OHare:2017yze}
C.~A.~J. O'Hare and A.~M. Green, \emph{{Axion astronomy with microwave cavity experiments}}, \href{https://doi.org/10.1103/PhysRevD.95.063017}{\emph{Phys. Rev. D} {\bfseries 95} (2017) 063017} [\href{https://arxiv.org/abs/1701.03118}{{\ttfamily 1701.03118}}].

\bibitem{Diehl:2023fuk}
J.~Diehl, J.~Knollm\"uller and O.~Schulz, \emph{{Bias-free estimation of signals on top of unknown backgrounds}}, \href{https://doi.org/10.1016/j.nima.2024.169259}{\emph{Nucl. Instrum. Meth. A} {\bfseries 1063} (2024) 169259} [\href{https://arxiv.org/abs/2306.17667}{{\ttfamily 2306.17667}}].

\bibitem{PhysRevD.42.3572}
M.~S. Turner, \emph{Periodic signatures for the detection of cosmic axions}, \href{https://doi.org/10.1103/PhysRevD.42.3572}{\emph{Phys. Rev. D} {\bfseries 42} (1990) 3572}.

\bibitem{2012ApJ...759..131B}
J.~{Bovy}, C.~{Allende Prieto}, T.~C. {Beers}, D.~{Bizyaev}, L.~N. {da Costa}, K.~{Cunha} et~al., \emph{{The Milky Way's Circular-velocity Curve between 4 and 14 kpc from APOGEE data}}, \href{https://doi.org/10.1088/0004-637X/759/2/131}{\emph{\apj} {\bfseries 759} (2012) 131} [\href{https://arxiv.org/abs/1209.0759}{{\ttfamily 1209.0759}}].

\bibitem{CAST:2024eil}
{\scshape CAST} collaboration, K.~Altenm\"uller et~al., \emph{{New Upper Limit on the Axion-Photon Coupling with an Extended CAST Run with a Xe-Based Micromegas Detector}}, \href{https://doi.org/10.1103/PhysRevLett.133.221005}{\emph{Phys. Rev. Lett.} {\bfseries 133} (2024) 221005} [\href{https://arxiv.org/abs/2406.16840}{{\ttfamily 2406.16840}}].

\bibitem{PhysRevLett.133.211002}
C.~A. Manzari, Y.~Park, B.~R. Safdi and I.~Savoray, \emph{Supernova axions convert to gamma rays in magnetic fields of progenitor stars}, \href{https://doi.org/10.1103/PhysRevLett.133.211002}{\emph{Phys. Rev. Lett.} {\bfseries 133} (2024) 211002}.

\bibitem{Dolan:2022kul}
M.~J. Dolan, F.~J. Hiskens and R.~R. Volkas, \emph{{Advancing globular cluster constraints on the axion-photon coupling}}, \href{https://doi.org/10.1088/1475-7516/2022/10/096}{\emph{JCAP} {\bfseries 10} (2022) 096} [\href{https://arxiv.org/abs/2207.03102}{{\ttfamily 2207.03102}}].

\bibitem{magnet_ieee}
A.~Torre and L.~Quettier, \emph{{Editorial of the \MADMAX{} Special Section}}, \href{https://doi.org/10.1109/TASC.2023.3309428}{\emph{IEEE TAS} {\bfseries 33} (2023) 1}.

\bibitem{open}
\MADMAX{} Collaboration, First search for axion dark matter with a \MADMAX{} prototype: Datasets (2025) \url{https://zenodo.org/records/15496833}.

\bibitem{4065916}
H.~A. Haus, W.~R. Atkinson, G.~M. Branch, W.~B. Davenport, W.~H. Fonger, W.~A. Harris et~al., \emph{Representation of noise in linear twoports}, \href{https://doi.org/10.1109/JRPROC.1960.287381}{\emph{Proceedings of the IRE} {\bfseries 48} (1960) 69}.

\end{thebibliography}\endgroup

\onecolumngrid
\clearpage

\setcounter{equation}{0}
\setcounter{figure}{0}
\setcounter{table}{0}
\setcounter{page}{1}
\makeatletter
\renewcommand{\theequation}{S\arabic{equation}}
\renewcommand{\thefigure}{S\arabic{figure}}
\renewcommand{\thepage}{S\arabic{page}}
\renewcommand{\thetable}{S\arabic{table}}

\begin{center}
\textbf{\large Supplemental Material for the Letter\\[0.5ex]
 First search for axion dark matter with a \MADMAXbf{} prototype}
\end{center}

\section{Coupling of axion field to TE$_{11}$ mode: Overlap formalism}
We implemented the dielectric haloscope principle in a closed cylindrical volume. 
The conducting boundaries of CB200 form a waveguide in which wave propagation is described by cylindrical solutions (modes) of the wave equation. To optimize the power coupling of the axion-induced field, we maximize the transverse overlap, $|\eta_{A}|$, between the uniform axion-induced field and the mode of propagation. The TE$_{11}$-mode maximizes power coupling and $|\eta_{A}|$ can be calculated as: 
\begin{equation}
	|\eta_{A}|^2 = \frac{|\int_A dA~\textbf{E}_{\mathrm{TE_{11}}}\cdot~\hat{y}|^2}{A \int_A dA~|\textbf{E}_{\mathrm{TE_{11}}}|^2} = 0.84,
\end{equation} 
with $A$ the transverse area of the booster and $\textbf{E}_{\mathrm{TE_{11}}}$ the electric field distribution of the TE$_{11}$ mode. This value can be interpreted as a scaling form-factor that reduces $\beta^2$. It is independent of frequency and applies for any ideal configuration of the booster implemented in cylindrical boundary conditions under the TE$_{11}$ mode. It is therefore implicitly included everywhere in our $\beta^2$ calculations.

The TE$_{11}$ cylindrical wave propagates along the main axis where the transverse field distribution is mainly unchanged due to symmetry. To simplify the analysis, it is common to decouple the propagation direction, treating the problem as 1D. This approach is advantageous because it requires relatively few well-defined parameters to determine $\beta^2$. Such a single mode booster model was already demonstrated to fit the complex reflection coefficient S11 obtained from the response of the booster \cite{Egge:2020hyo}. 

\section{Simulation of the electromagnetic (EM) booster response}
A large booster diameter, much larger than the wavelength, 
is required to increase sensitivity. This results in a high number of parasitic modes. 
Mode-crossings between the desired and parasitic modes alter the field distribution and degrade the axion signal. While the single mode booster model can be used for properly describing the booster mode, finite element simulations are needed to understand the complex over-moded behavior that includes the parasitic modes.
Such simulations, including the taper, the Rexolite\textsuperscript{TM} lens and the booster geometries implemented in COMSOL\textsuperscript{TM} using the corresponding material constants, were performed.
\begin{figure}[b!]
	\centering
	\includegraphics[width=1\textwidth]{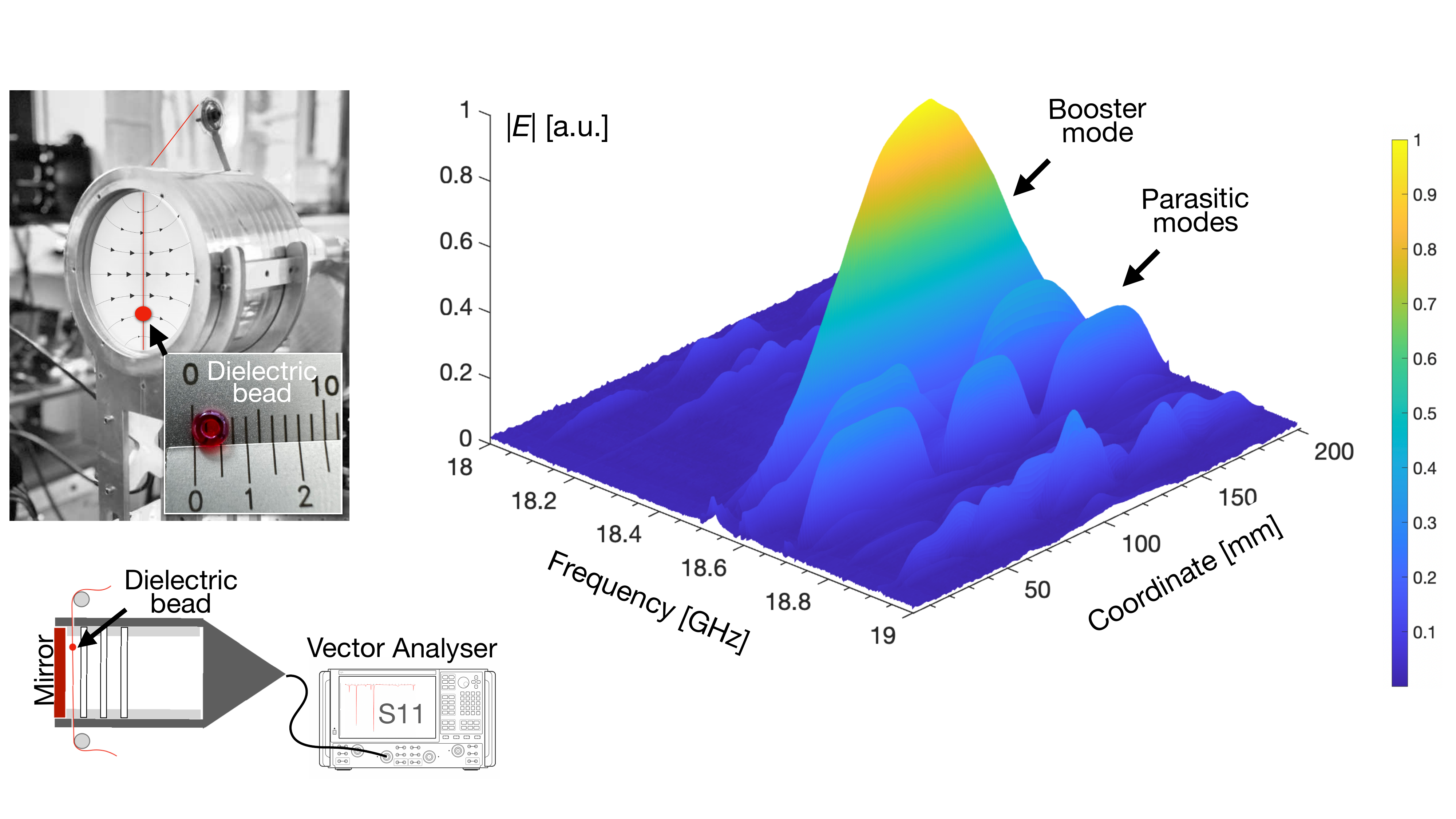}
	\caption{Details of the bead-pull measurement used to identify the booster mode. Left: Schematic of the bead-pull implementation and a photo of the opened setup. The expected E-field of the desired TE$_{11}$ mode and a sketch of the bead are overlayed on the photo. The E-field strength is resolved along the profile indicated by the red line near the mirror surface. Right:  The measured electric field distribution is shown as a function of frequency along the diameter of the booster cavity, with the center located at the coordinate 100 mm. The field profile of the booster mode exhibits a well-defined single peak, consistent with the expected the TE$_{11}$ mode.
		}
	\label{fig:bead_pull}
\end{figure}

Different booster configurations are modeled by adapting the spacing between the disks and the mirror and the distance to the taper aperture, L.
Losses are accounted for by the loss tangent of the dielectric lens and disks and the finite conductivity of all metallic walls.
The attenuation of the lens is estimated to be < 0.1 dB. 
The simulations show that  with increasing L modes originating from the taper are shifted in frequency. 
In contrast to this, the booster mode is unchanged compared to the rest of the spectrum. Modifying $L$ can therefore be used to avoid crossings with parasitic modes. 

The key outcomes of the simulations are: 
\begin{enumerate}
    \item  We have established a procedure to tune the frequency of the booster mode and avoid mode crossings.
    \item We have verified that with sufficient frequency separation between the parasitic resonances and the desired booster mode, the field distribution and booster line remain largely unaffected, allowing the booster mode to be well fitted by the single mode booster model.
    \item We can identify parasitic modes in regions of the spectrum near the booster mode. 
\end{enumerate}
The outlined strategy has been used throughout the experiment to tune the frequency of the booster mode at multiple frequencies in the 18 to 21 GHz\,range.

Misalignment can introduce asymmetry in the E-field, thereby affecting the E-field of the booster mode. 
In the current resonant configuration the strongest effect arises from the strong field between reflective mirror and first disk, which makes the mode sensitive to a mirror tilt. Our measurements also show sensitivity to mirror tilt during the tuning process.  COMSOL\textsuperscript{TM} simulations have shown that the typical tilts seen in our setup have only a perturbative effect on the field distribution of the booster mode. This leads to relatively small boost factor degradation, even for relatively large tilts of up to 0.2$^{\circ}$ corresponding to a $\approx\,$350\,$\mu$m offset at the mirror edge, which shift the booster mode frequency by around 15\,MHz and decrease the boost factor by up to 15\%. These effects are well within the range accounted for by the single mode booster model used to fit the physics run power spectra and are included in the uncertainty budget on boost factor determination \autoref{tab:systematics}.

\section{Verification of the booster mode}
Direct E-field measurements using the bead-pull method \cite{Egge:2023cos} were used to
confirm that the E-field of the booster mode follows the expected shape of a TE$_{11}$ mode.
The setup used is reported in figure \autoref{fig:bead_pull} (left). 
The measurement is conducted along a single axis, perpendicular to the polarization, resulting in a radial 1D field profile along the entire diameter of the booster. In figure \autoref{fig:bead_pull} (right) we report the measured electric field as a function of frequency. The field profile of the booster mode shows a well-defined single peak as expected for the case of a TE$_{11}$ mode. Parasitic modes are easily identifiable by their higher-order variations in the field profile. Consistent 
 behavior of the field distribution has been reproduced throughout the preparation of all booster configurations.

\begin{figure}[t!]
    \centering
    \includegraphics[width=0.9\textwidth]{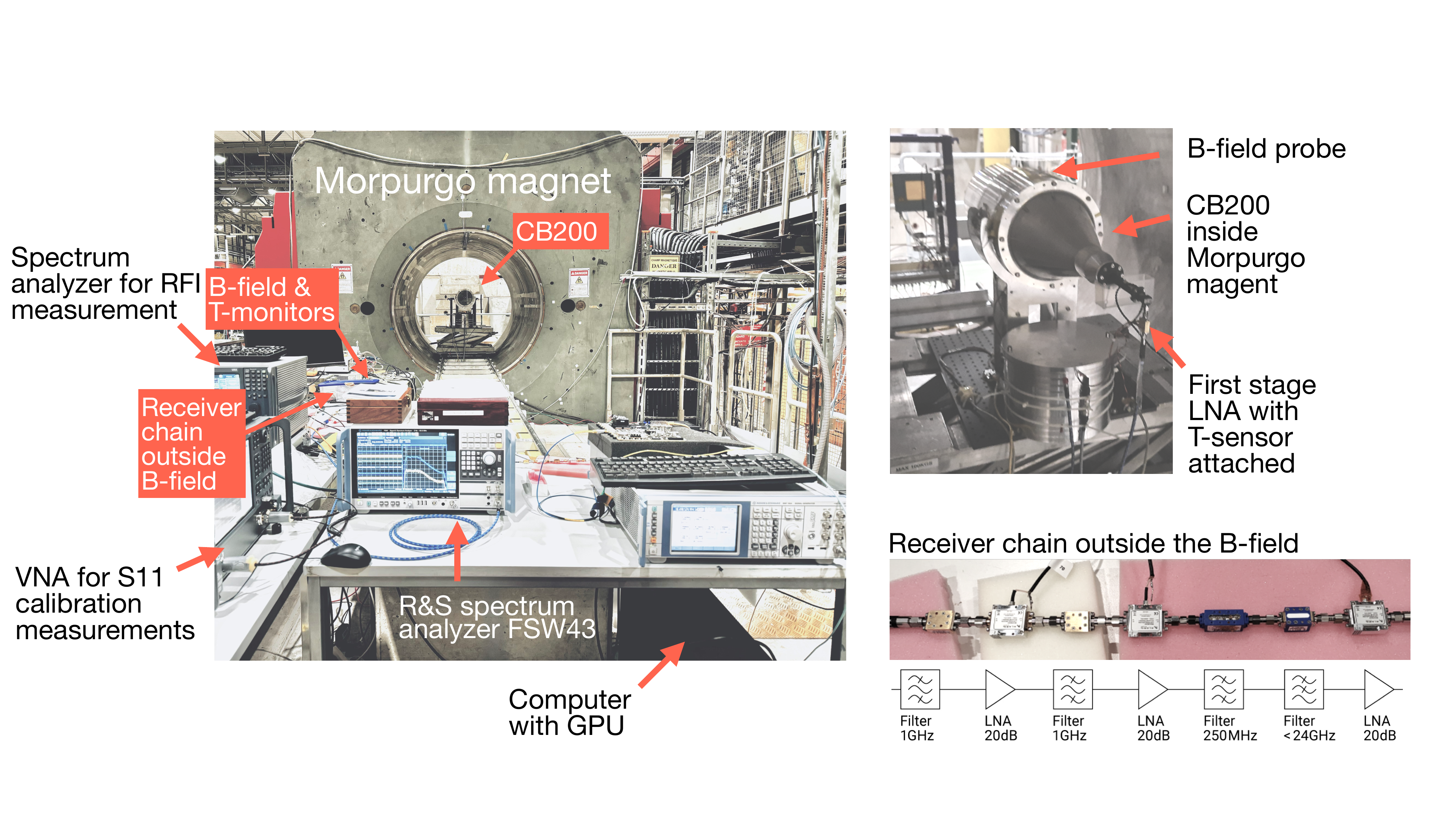}
    \caption{Left and top right: Picture of CB200 inside the Morpurgo magnet and equipment for data taking and data monitoring. Bottom right: Receiver chain components outside the B-field.}
    \label{fig:setup_labelled}
\end{figure}

\section{Further information on data taking}
\begin{table*}[b!]
    \centering
     \caption{Overview over datasets taken.}
    \begin{tabular}{c|c|c|c|c}
        $\beta^2$ Peak & Frequency& Median & Number of  & Corresponding \\        
          maximum      & region   & B-field& 14.75 min. files & integration time\\
\hline
        [GHz]  & [GHz]  &  [T] &       & [h] \\
\hline
        18.531 & 18.513-18.575 & 1.58 & 54  & 13.28  \\
        18.543 & 18.513-18.575 & 1.01 & 295 & 72.53  \\
        18.557 & 18.513-18.575 & 1.01 & 636 & 156.38  \\
        19.196 & 19.177-19.230 & 1.27 & 163 & 40.08  \\
        19.215 & 19.177-19.230 & 1.58 & 245 & 60.24 \\ 
    \end{tabular}
   
    \label{tab:data_taking}
\end{table*}

\begin{figure*}[t!]
    \centering
    \includegraphics[width=0.7\linewidth]{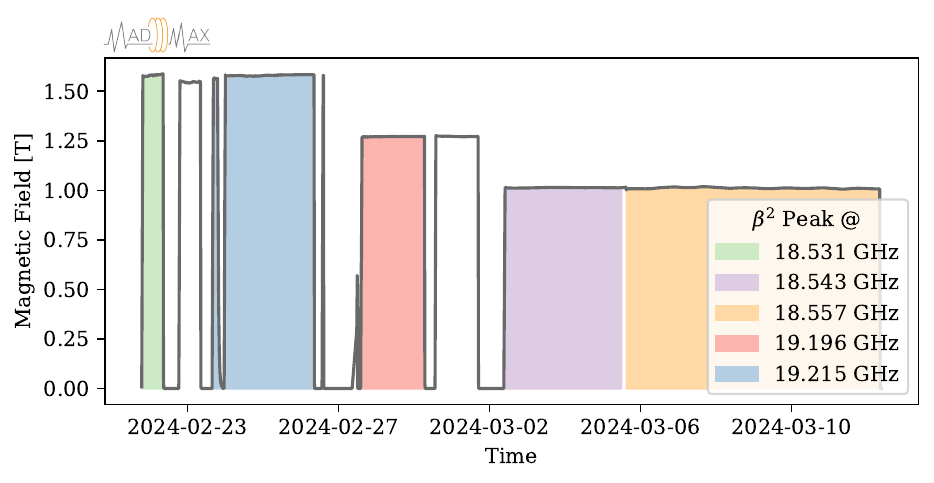}
    \caption{
    Time evolution of the available magnetic field strength. Individual physics-runs are denoted with various colors. Due to issues encountered with the cryogenic system of the magnet after first measurements at $\SI{1.6}{\tesla}$ the field strength had to be reduced to $\SI{1.3}{\tesla}$ and ultimately to $\SI{1.0}{\tesla}$. Two time periods have not been used for physics measurements (white). For the first time period test measurements were conducted with a different receiver system. During the second time period the booster peak frequency displayed significant variation over time.}
    \label{fig:data_taking} 
\end{figure*}

The experimental setup during data taking is shown in \autoref{fig:setup_labelled}.
The B-field data referred to in the main text and shown in \autoref{fig:data_taking} were taken using a Lakeshore model 475 DSP gaussmeter.
The B-field probe was attached to the top of the CB200 cylinder, $\approx$\,11\,cm above the center of the magnet bore. CB200 was placed within the magnet with the mirror  centered within the length of the magnet bore to 1\,cm precision. 
3D measurements of the magnetic field performed by the CERN magnet group demonstrated that the uncertainty on the B-field strength is 2\%, which has negligible impact on the  $|g_{a\gamma}|$ limit.

\autoref{tab:data_taking} summarizes the data taken during the individual runs, specifying the boost factor peak maximum, the frequency range in which data has been taken, the median B-field during the run, the number of individual 14.75 minute files and the corresponding integration time of the overall run. 
The measured B-field as a function of time is shown in \autoref{fig:data_taking}.
The median change of the B-field in between individual 14.75 minute files was 0.15\,mT leading to negligible uncertainty on the limit.

\section{Y-factor calibration}

In order to calibrate the gain of the receiver chain a Y-factor calibration was performed  using a noise source Keysight 346C directly connected to an attenuator of 50\,$\Omega$ impedance with attenuation A=$30$\,dB. At room temperature and for our frequency range of interest we work with the Rayleigh-Jeans approximation of Planck's law for the noise power $P_N$:
\begin{equation}
    P_N = k_B B T
\end{equation}
with $k_B$ the Boltzmann constant, $B$ the resolution bandwidth of the receiver and $T$ the equivalent noise temperature. The warm and cold measurements of the Y-factor calibration correspond to the noise source being switched on and off. 
For both cases, the expected noise power has two contributions: one is coming from 
the equivalent noise temperature of the amplifier
the other from the attenuator with noise source attached. 
With the noise source off, the attenuator emits noise equivalent to a resistor at room temperature with $T_0=290\,$K.
For the measurement with the noise source on, the noise equivalent power of the attenuator is given by the excess noise ratio (ENR) of the noise source, attenuated by A.

In addition, the amplifier is not matched to the $50$ $\Omega$ impedance of the amplifier which we also take into account. Therefore:
\begin{equation}
\begin{split}
    P_{N,{\mathrm{off}}} = k_B G B T_0 (1-\mid \Gamma_R \mid ^2) + k_B G B T_e  \\
    P_{N,{\mathrm{on}}} = k_B G B T_0\cdot (1 + 10^{(\mathrm{ENR-A})/10})(1-\mid \Gamma_R \mid ^2) + k_B G B T_e
\end{split}
\end{equation}
where $G$ is the gain of the amplifier, $\Gamma_R$ is the reflection coefficient due to the mentioned mismatch between amplifier and attenuator, $T_e$ is the equivalent noise temperature of the amplifier and an ENR$\,=\,15.346$ has been taken according to the specifications provided by the manufacturer. 

No significant change in LNA gain as function of B-field was observed.

\section{Booster noise model}
The noise model of booster and receiver chain consists of the noise emitted by the first stage amplifier and the booster itself.
For the first stage amplifier, a noise model as per \cite{4065916}  
is used, simulating its noise by a correlated voltage and current source with the parameters extracted from measurements with calibrated standards (short, open, load) connected to the amplifier.

\begin{figure*}[t!]
    \centering
    \includegraphics[width=0.6\textwidth]{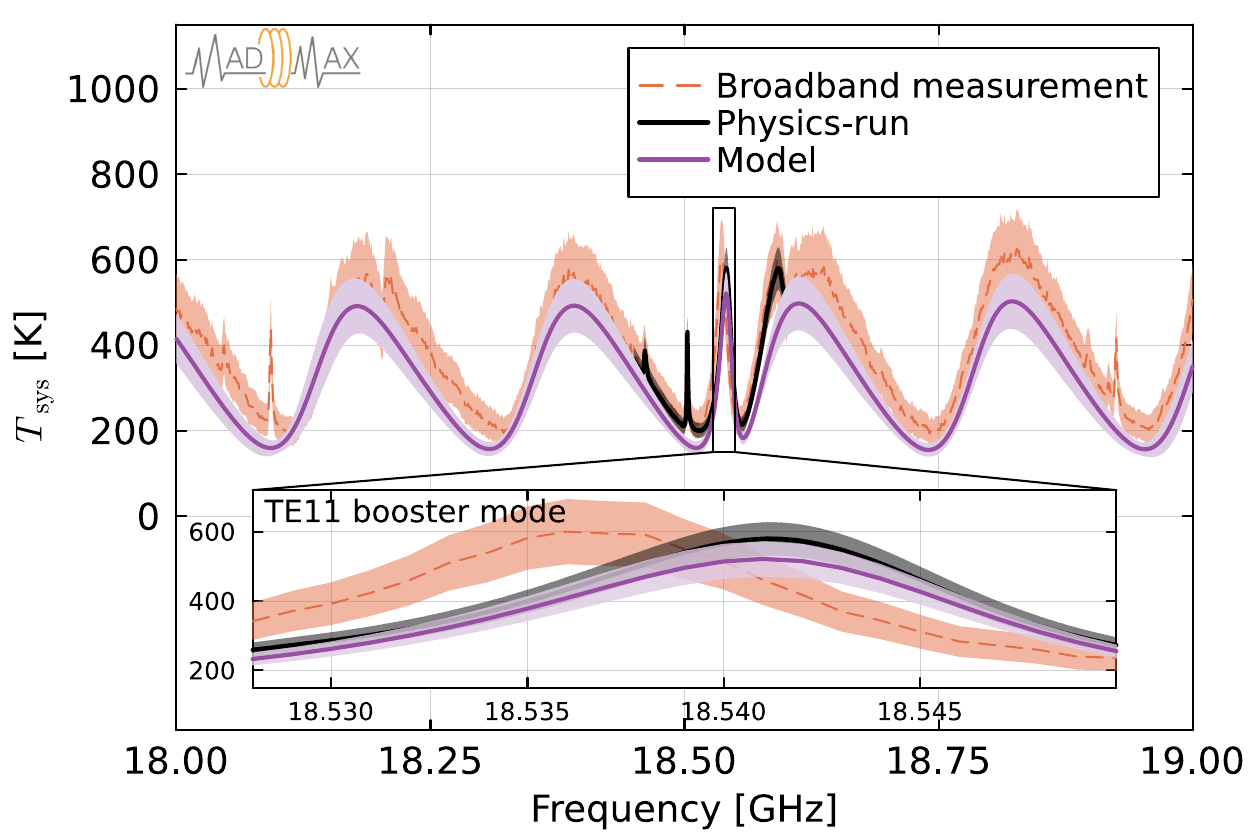}
    \caption{
    Exemplary system temperature spectra ($T_\mathrm{sys}$) obtained from broadband measurements (dashed red) and
    from an axion physics-run (solid black). For the latter, only a reduced frequency range around the TE$_{11}$ mode resonance is recorded. The $T_\mathrm{sys}$ spectrum obtained from the booster model (solid magenta) is superimposed with the band around the line indicating the systematic uncertainties. 
    For the physics-run power spectrum, a zoom to the TE$_{11}$ booster mode region is shown in the inset and a good agreement  with the model is observed. 
       }
    \label{fig:booster_noise}
\end{figure*}%

The booster noise is modeled by gray body radiation with the emissivity parameter corresponding to the losses which can be determined  by fitting the model to the measured reflectivity. In the the ADS model, booster and amplifier are connected by a transmission line of length $d$. Noise from both sources is reflected by the mismatched amplifier and the booster causing interference effects seen in \autoref{fig:booster_noise}. 
Simulating only amplifier or only booster emissions shows that the noise spectrum is dominated by the amplifier noise oscillating between ~100 and 500 K. The oscillation in noise temperature is due to a standing wave between amplifier and booster.
For the broadband measurements, the model describes well the standing wave pattern induced by the amplifier noise ($\sim$\,230\,K, see text) inside the booster, but not the fine structures indicating the limitation of the one-dimensional model.
The shift of the TE$_{11}$ resonance peak frequency between broadband and physics spectrum can be attributed to environmental effects, specifically the tuning mechanism being affected during magnet ramp up/down -- the shift shown is the biggest observed in all datasets. To match the resonance position of the physics-run measurement after the system stabilized, the distance between mirror and closest disk used in the booster model was adjusted, requiring $\mathcal{O}(\si{\micro\m})$ changes from the previously determined parameters.
The value for $d$ is extracted from a fit to the calibration broadband measurement.
The change in boost factor due to the connected amplifier is quantified by the factor 
\begin{equation}    
F_{RC} = \frac{1 - |\Gamma_{\mathrm{LNA}}|^2}{|1 - e^{-2i\omega d}\Gamma_b\Gamma_{\mathrm{LNA}}|^2} 
\end{equation}
with the boosters and LNA reflectivities $\Gamma_b$ and $\Gamma_{\mathrm{\mathrm{LNA}}}$.

\begin{figure*}[b!]
    \centering
    \includegraphics[width=0.8\linewidth]{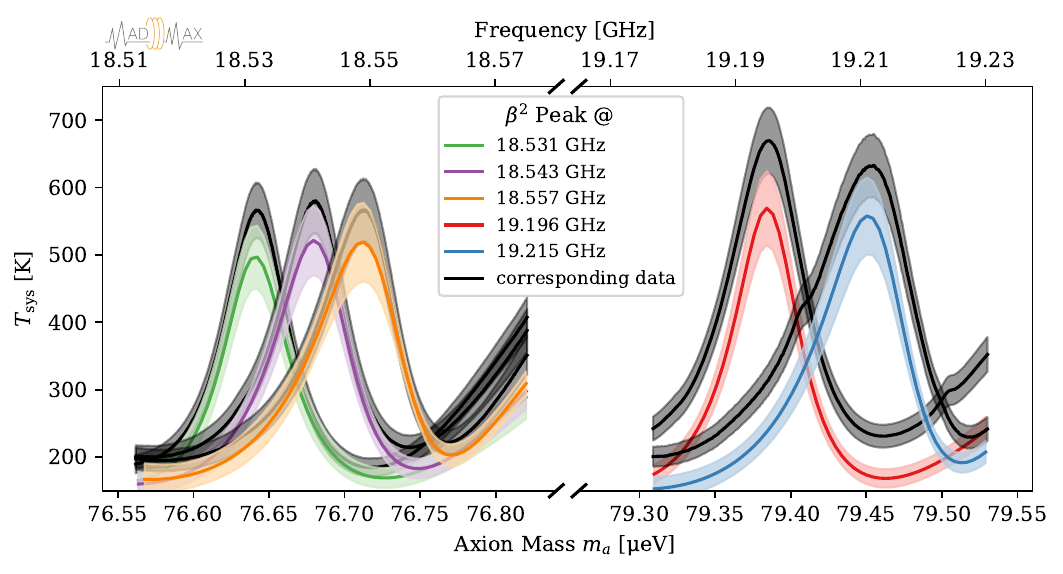}
    \caption{
       Simulation and data for each physics-run's power spectrum around the TE$_{11}$ booster mode. 
       }
    \label{fig:power_spectra}
\end{figure*}

The booster noise has a significant contribution of $\approx$\,150\,-\,200\,K at the boosters TE$_{11}$ resonance frequency. It results from the following components: 
\begin{enumerate}
    \item waveguide, taper and lens. These are modeled by one single attenuation contributing with roughly 10 K noise temperature,
    \item the disks, that, from the effective loss determined in the fitting procedure, have a noise temperature of $\approx$\,0.03\,K each and
    \item  the mirror with a noise temperature of $\approx$\,0.3\,K, determined by its absorption calculated from its conductivity. 
\end{enumerate}
Outside the boosters TE$_{11}$ resonance, the simulated booster noise oscillates around $\approx$\,10\,K, matching the incoherent sum of those components. On resonance, as a very rough estimate, the $\approx$\,0.3\,K equivalent noise temperature of the mirror can be multiplied by the boost factor of $\mathcal{O}(10^3)$ to get to the maximum noise contribution of $\mathcal{O}(10^2)$ K of the booster.

Simulates and measured system noise temperature for all five physics runs are shown in \autoref{fig:power_spectra}. The model reproduces the frequency position of the resonance peaks. System temperature is consistently slightly underestimated,
    which does not affect the boost factor determination, as only the frequency dependence of $T_\mathrm{sys}$ is used in the analysis.

\section{Further information}
Data taking and analysis involve many different frequency bands. They are summarized and commented in \autoref{tab:R-datasets}.

\begin{figure*}[ht]
    \centering
    \includegraphics[width=0.8\linewidth]{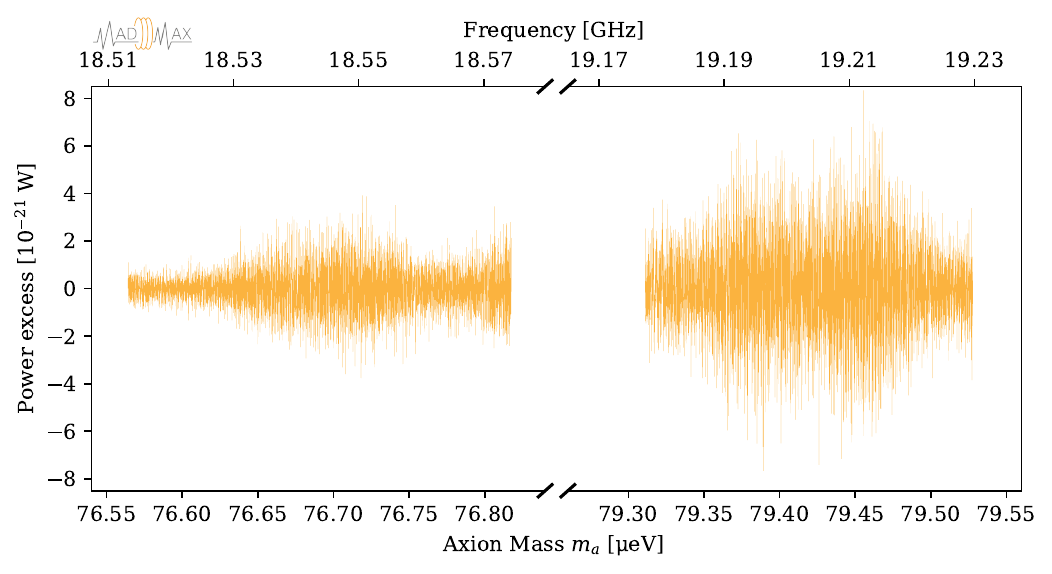}
    \caption{Observed cross-correlated power excess as a function
of frequency.
    }
    \label{fig:excess_power_spectra}
\end{figure*}

\autoref{fig:excess_power_spectra} Shows the power excess in each individual frequency bin after cross correlation with the expected axion signal shape.

\begin{figure*}[ht]
    \centering
    \includegraphics[width=0.8\linewidth]{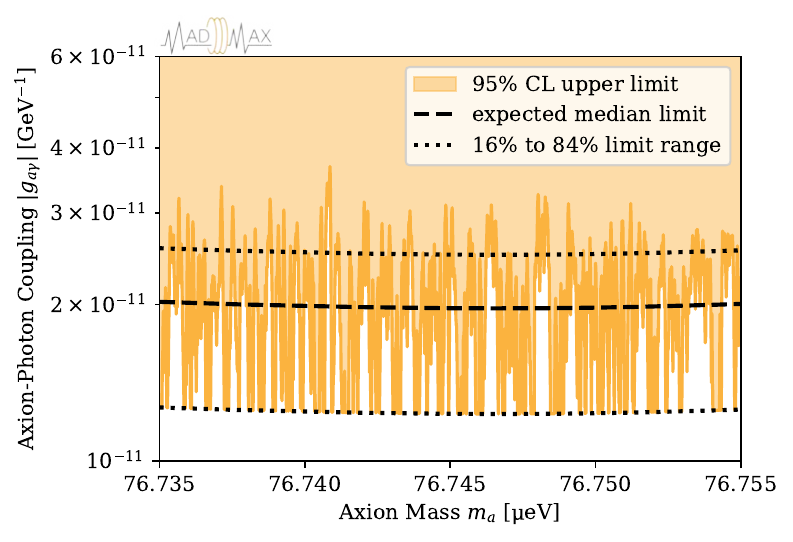}
    \caption{95\% CL exclusion limit around the most sensitive region for configuration~1. 
        }
        \label{fig:limit_zoom}
\end{figure*}

\autoref{fig:limit_zoom} shows a zoom of the limit plot at the axion mass range with highest sensitivity.
The bin to bin fluctuation of the limits and the truncation below the 16\% quantile are clearly visible, as are correlations between adjacent frequency bins. The latter arise due to cross-correlation with the axion line-shape as mentioned in the text.

\begin{table}[t!]
   \caption{
    Summary of frequency bands relevant in the data taking, calibration and analysis procedure.}
    \vspace{2mm}
    \centering
    \begin{tabular}{c | c | c}
    \toprule
               & Frequency   & Description \\
\hline
\hline
    \makecell{Raw data bins\\ resolution bandwidth} & 9 Hz & \makecell{Given by inverse of coherent integration time of DAQ, i.e. the \\ length of the individual time domain signal traces with \\ 110\,ms length which are FFT-ed.} \\
  \hline
   \makecell{Analysis data  bins\\  resolution bandwidth} & 0.9\,kHz & \makecell{Chosen such that frequency resolution is good enough to resolve a \\ potential axion signal-width.} \\
  \hline
   
   \makecell{FWHM of Axion, \\ signal width} & 20 kHz & Given by DM velocity dispersion of galactic DM halo  \\
  \hline
  
   \makecell{Boost factor \\ band width} & 10 MHz & \makecell{Depends on booster configuration. In this case with 3 disks \\ we are using resonant configuration determined by distance \\ between mirror and first disk corresponding to roughly \\ $\lambda$/2 of resonant frequency.}\\
  \hline
   
   \makecell{Receiver bandwidth \\ physics runs} & 250 MHz & \makecell{Adapted to boost factor width and scan width. Determined by pass band\\ filters in front of receiver system and by its settings.}\\
  \hline
   
   \makecell{Spectrum analyzer bandwidth \\ during calibration} & 1 GHz & \makecell{Determined by peak to peak fluctuation length (in frequency domain) \\ of Fabry-Perot oscillation pattern due to LNA noise induced \\ standing wave between LNA and booster.} \\
  \hline

 \bottomrule
    \end{tabular}
\label{tab:R-datasets}    
\end{table}

\clearpage

\end{document}